\documentclass[instruments,article,accept,pdftex,moreauthors]{Definitions/mdpi}

\firstpage{1} 
\pubvolume{8}
\issuenum{4}
\articlenumber{48}
\pubyear{2024}
\copyrightyear{2024}
\externaleditor{Academic Editor: Pasquale Arpaia}
\datereceived{23 September 2024} 
\daterevised{17 October 2024} 
\dateaccepted{1 November 2024} 
\datepublished{4 November 2024} 
\hreflink{https://doi.org/10.3390/\linebreak instruments8040048} 

\usepackage{xcolor}
\usepackage[labelformat=simple]{subcaption}
\DeclareCaptionLabelFormat{subcaptionlabel}{\normalfont(\textbf{#2}\normalfont)}
\usepackage{graphicx}
\usepackage{float}
\usepackage{siunitx} 

\Title{Improvement and Characterisation of the ArCLight {Large-Area Dielectric Light Detector for Liquid-Argon} 
	 Time Projection~Chambers}

\TitleCitation{Improvement and Characterisation of the ArCLight {Large-Area} Dielectric Light Detector for {Liquid-Argon }Time Projection Chambers}

\Author{Jonas Bürgi \orcidH{}, Livio Calivers \orcidA{}, Richard Diurba \orcidE{}, Fabian Frieden, Anja Gauch \orcidG{}, Laura Francesca Iacob $^{\dagger}$, Igor~Kreslo $^{\ddagger}$\orcidD{}, Jan Kunzmann \orcidF{}, Saba Parsa \orcidC{} and Michele Weber *\orcidB{}} 

\AuthorNames{Jonas Bürgi, Livio Calivers, Richard Diurba, Fabian Frieden, Anja Gauch, Laura Francesca Iacob, Igor Kreslo, Jan Kunzmann, Saba Parsa and Michele Weber}

\AuthorCitation{Bürgi, J.; Calivers, L.; Diurba, R.; Frieden, F.; Gauch, A.; Iacob, L.F.; Kreslo, I.; Kunzmann, J.; Parsa, S.; Weber, M.}

\address[1]{%
{Laboratory} for High Energy Physics, Albert Einstein Center for Fundamental Physics, Universität Bern, 3012~Bern,~Switzerland 

}

\corres{\hangafter=1 \hangindent=1.05em \hspace{-0.82em}Correspondence: michele.weber@unibe.ch}

\firstnote{\hangafter=1 \hangindent=1.05em \hspace{-0.82em}{Current address:} Università degli Studi di Perugia, 06124 Perugia, Italy.} 
\secondnote{\hangafter=1 \hangindent=1.05em \hspace{-0.82em}{Current address:} Moscow Institute of Physics and Technology, {141701 Dolgoprudny}, Russia.} 

\abstract{The detection of scintillation light in noble-liquid detectors is necessary for identifying neutrino interaction candidates from beam, astrophysical, or solar sources. Large monolithic detectors typically have highly efficient light sensors, like photomultipliers, mounted outside their electric field. This option is not available for modular detectors that wish to maximize their active volume. The ArgonCube light readout system detectors (ArCLights) are large-area thin-wavelength-shifting (WLS) panels that can operate in highly proximate modular detectors and within the electric field. The WLS plastic forming the bulk structure of the ArCLight has Tetraphenyl Butadiene (TPB) and sheets of dichroic mirror layered across its surface. It is coupled to a set of six silicon photomultipliers (SiPMs). This publication compares TPB coating techniques for large surface areas and describes quality control methods for large-scale production.}


\begin{document}


\section{Introduction}
Long-baseline neutrino detectors rely on large-area photon detectors such as photomultipliers to detect Cherenkov or scintillation light from neutrino interactions. Examples of such detectors are the Hyper-Kamiokande~\cite{HK} and the Deep Underground Neutrino Experiment (DUNE)~\cite{DUNETDRVol1} Far Detector modules. These detectors expect to measure no more than a few neutrinos a day. Nevertheless, future neutrino beams, like Fermilab's PIP-II beam, will provide $\mathcal{O}$($100$) of neutrino interactions for the liquid-argon near detector~\cite{duneNDCDR} per beam spill. Multiple light flashes belonging to different neutrino interactions could coincide within the long drift time of the charge signals and lead to an overlap of signals from different interactions. Modularisation of the detector can isolate the distinct scintillation light signals induced by the many neutrino scatterings per spill in optically segmented volumes. However, the~light detectors must have a small volume to minimize dead space between modules~\cite{Amsler:1993255}.

The ArgonCube light readout system (ArCLight) addresses the need for dielectric, large-area, small-footprint photon detectors. The~dielectric design of the ArCLight structure allows it to be placed directly in a detector's high-voltage drift field. Accordingly, DUNE has selected ArCLight for the DUNE liquid-argon near detector (ND-LAr).
Inspired by the ARAPUCA light trap~\cite{Machado:2016jqe}, ArCLight follows a similar principle: light is trapped and travels through a wavelength-shifting (WLS) layer towards an array of six SiPMs coupled along one face of the WLS plate~\cite{instruments2010003}. 

This publication outlines the production methods used to produce ArCLights in large quantities and the quality control methods utilised to ensure that high-performance ArCLights can operate in modular detectors. The~ND-LAr detector comprises 35 modules, each measuring approximately \SI{100}{\centi\meter} in width, \SI{100}{\centi\meter} in length, and~\SI{300}{\centi\meter} in height. Each module contains two time projection chambers with an anode plane on the opposing walls. Fifty percent of the area of the side walls is covered by ArCLights. The~remainder of each side wall is filled by an alternate light trap design based on wavelength-shifting fibers that span the surface of a polycarbonate back panel. The~fiber-based modules are called Light Collection Modules (LCMs) and provide an increased photon detection efficiency~\cite{LCMs}. 

The ArCLights to be deployed in the DUNE ND-LAr are \SI{30}{\centi\meter} wide, \SI{50}{\centi\meter} long and \SI{1}{\centi\meter} thick. For~the studies in this publication, ArCLights with a reduced length of \SI{28}{\centi\meter} were used. These were constructed for an ND-LAr demonstrator detector, the~ArgonCube 2 $\times$ 2 Demonstrator~\cite{duneNDCDR}. The~ArgonCube 2 $\times$ 2 Demonstrator is a modular liquid-argon detector containing four reduced-size prototypes of DUNE ND-LAr modules assembled in a two-by-two~arrangement.

In Section~\ref{sec:overview}, an~overview of the ArCLight technology as a light trap for vacuum ultraviolet (VUV) photons is provided. Section~\ref{sec:TPBCompare} describes two different Tetraphenyl Butadiene (TPB) coating techniques. Section~\ref{sec:production} describes the manufacturing process for ArCLights, and~Section~\ref{sec:qaqc} details the analysis performed on the quality control test data of the produced~ArCLights.

\section{Overview of ArCLight~Concept}\label{sec:overview}
Liquid argon emits scintillation light in a narrow band centered around \SI{128}{\nano\meter} with different decay times ranging from a few \SI{}{\nano\second} to \SI{}{\micro\second} when excited by a passing-through charged particle. The~time profile consists of a fast and slow component due to the decay of singlet and triplet excited states of the argon dimers~\cite{hitachi1983effect}. 
This light is collected by large-area light traps, which are read out by photosensitive devices. Given that VUV photons quickly become absorbed in materials commonly used in light traps, wavelength-shifting techniques are required to convert them to a suitable wavelength for propagation inside the bulk~material.

\begin{figure}[H]
 \hspace{-6pt}   \includegraphics[width=0.8\textwidth]{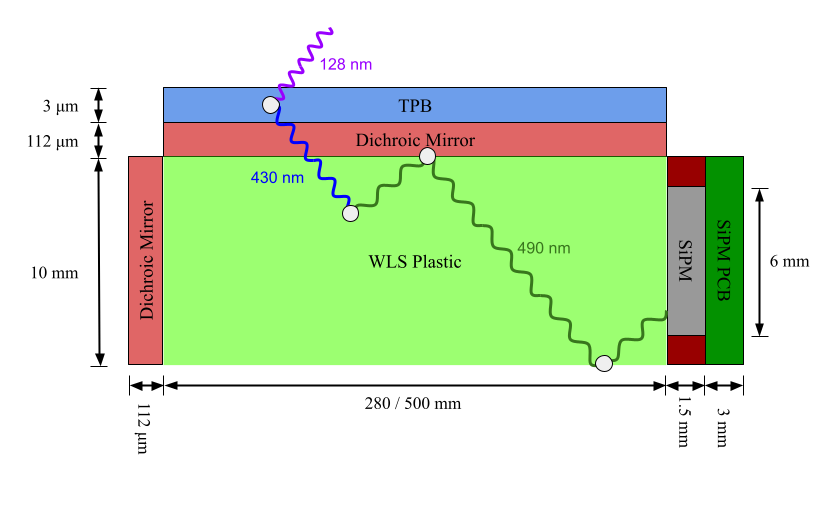}\vspace{-16pt}
    \caption{{Diagram} of an ArCLight with an example of a VUV photon travelling through the TPB and shifting to blue (peak at \SI{430}{\nano\meter}) to then pass into the dichroic mirror. The~photon shifts to green (peak at \SI{490}{\nano\meter}) inside the WLS plastic. The~green photons travelling in the plastic are eventually detected by the SiPM.}
    \label{fig:Arclight_layers}
\end{figure}

The design of the ArCLight incorporates two stages of wavelength-shifting. Figure~\ref{fig:Arclight_layers} shows a conceptual diagram of the~ArCLight.
The bulk material of the ArCLight consists of a \SI{10}{\milli\meter} thick plate of WLS material EJ280, produced by Eljen Technology, that is covered by a \SI{0.112}{\milli\meter} thick dichroic mirror (DF-PA Chill foil), produced by 3M. The~foil is coated with a layer of Tetraphenyl Butadiene (TPB) of approximately \SI{3}{\micro\meter} thickness to shift the VUV into blue (peak at \SI{430}{\nano\meter}) photons. \vspace{-12pt}

The~photons enter the WLS plate and are shifted into the green (peak at \SI{490}{\nano\meter}). Due to the dichroic mirror, the~photons are trapped inside the structure. The~dichroic mirror covers three narrow edges of the WLS plate to improve the trapping efficiency. On~the uncovered readout edge of the WLS plate, six Hamamatsu S13360-6050CS SiPMs with a \qtyproduct{6 x 6}{\milli\meter} sensitive area are mounted. The~SiPMs are spaced with up to a 10~mm deviation in spacing to satisfy mechanical constraints associated with the interface {of} 
the TPC structure.

One particular requirement for the ArCLight design is that it can be placed along the drift electric field of the TPC; therefore, the light trap is made of dielectric materials. The~readout edge with SiPMs and electronics is located at the anode plane of the TPC and, therefore, is not subject to a varying~potential.

\section{Comparison of TPB Coating~Techniques}\label{sec:TPBCompare}

\begin{figure}[H]
    \includegraphics[trim={0cm 0cm 0cm 0cm},clip,width=0.45\textwidth]{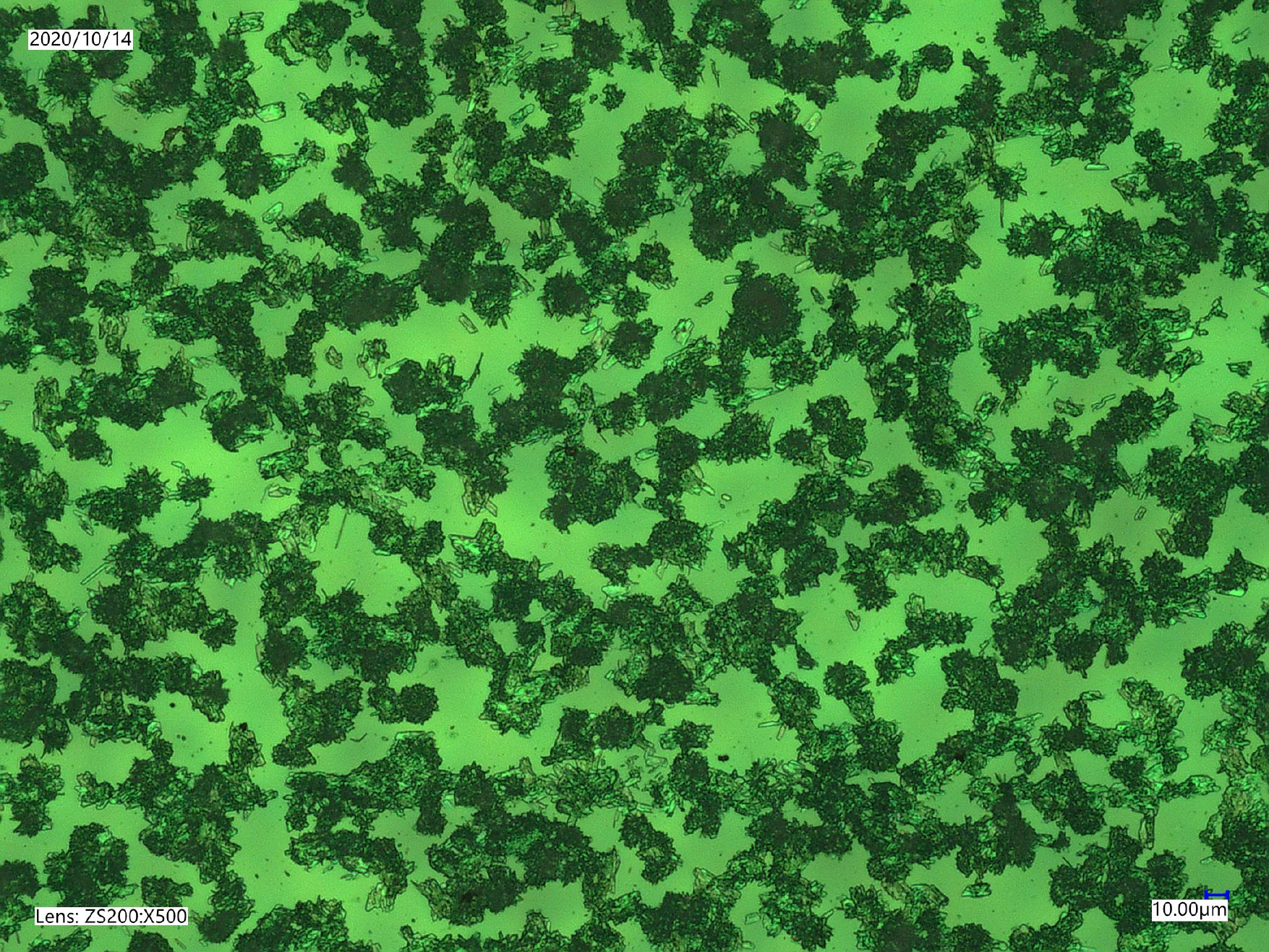}
    \includegraphics[trim={0cm 0cm 0cm 0cm},clip,width=0.45\textwidth]{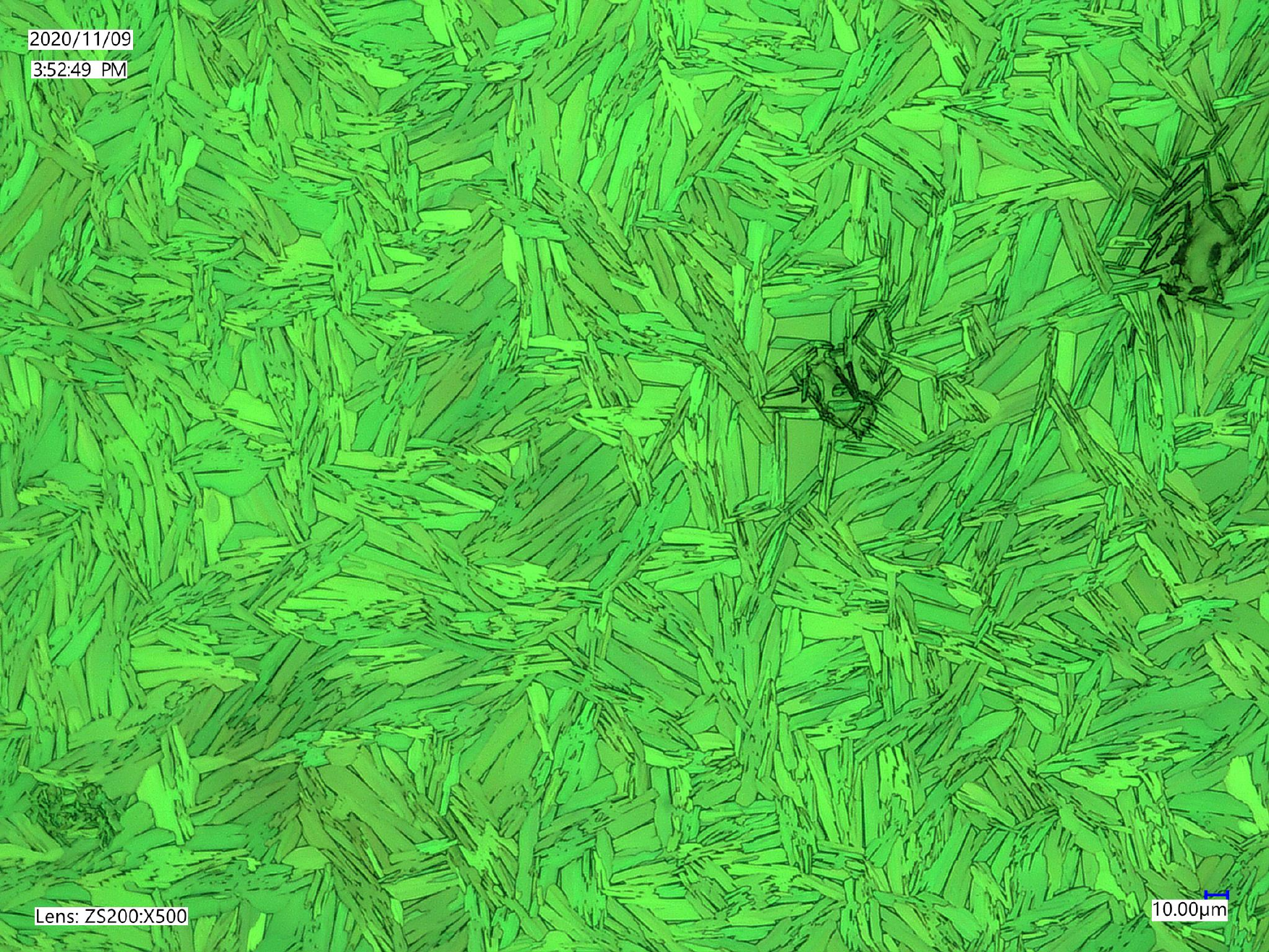}
    \caption[TPB layer structure for different coating techniques]{Microscopic images of the TPB layer achieved with airbrush (\textbf{left}) and evaporation deposition (\textbf{right}). Both images are magnified by a factor of 500.}
    \label{fig:tpb_microscope}
\end{figure}

A key stage in the production of ArCLights is the application of its TPB coating. In~early prototypes of the ArCLight, TPB was deposited by airbrush.
For this purpose, TPB was dissolved in toluene with polystyrene, then sprayed directly onto the dichroic mirror.
This method leads to a structure of TPB crystals embedded in polystyrene. It improves the robustness of the TPB layer,  helping to avoid dissolution in liquid argon over years of operation, as~has been observed previously~\cite{Asaadi:2018ixs}. 
However, inspecting the TPB layer under an optical microscope shows that the coverage factor of TPB is relatively poor using the airbrush method. Figure~\ref{fig:tpb_microscope} shows the microscopic view of the TPB~coverage.

As stated in~\cite{Benson2018}, vacuum evaporation deposition can achieve a higher TPB coverage. In~evaporation deposition, the~substrate coating, TPB, is vaporised in vacuum using a heat source. The~flux of the vaporised substrate eventually hits the target surface, where it condensates. An~evaporation chamber designed and optimised for coating the large surface of an ArCLight was developed and~tested.

The resulting crystal layer is shown in Figure~\ref{fig:tpb_microscope}. The~evaporation method, described in Section~\ref{sec:production}, creates a much more uniform crystal layer with an approximately full surface coverage. The~two coating methods were compared in a TPC setup using two identical ArCLights with different TPB layers. By~comparing the expected and measured light yield from cosmic muon tracks in the TPC, the~photon detection efficiency (PDE) was extracted. The~evaporation coating showed an improvement in efficiency by a factor of two compared to the airbrush method~\cite{Calivers:2886536}.

\section{Production of~ArCLight } \label{sec:production} 
\begin{figure}[H]
    \includegraphics[width=0.6\textwidth]{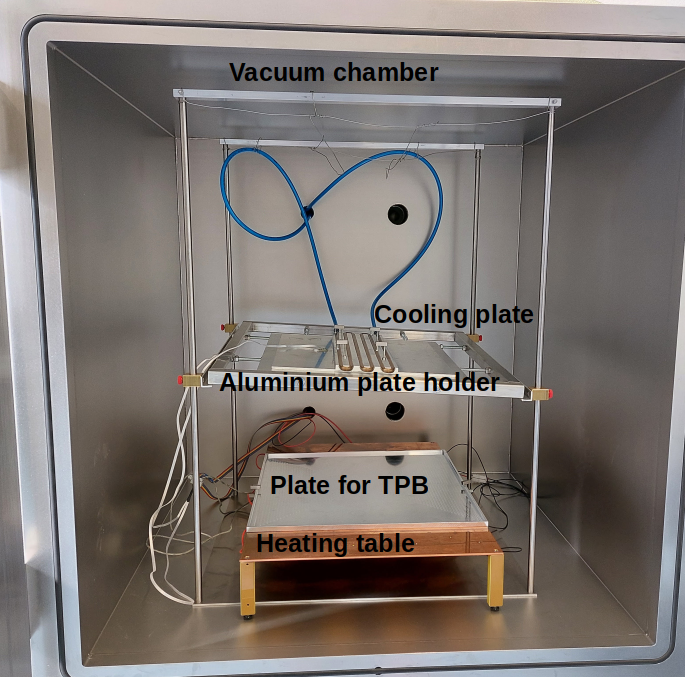}
    \caption{Image of the ArCLight TPB coating chamber. In~the vacuum chamber, the~TPB support plate is attached to the heating table. The~aluminium plate holder, on~which the cooling plate rests, hangs above~it.}
    \label{fig:coating_chamber} 
\end{figure}

The TPB coating is performed in a vacuum chamber from Pfeiffer Vacuum, KBH DN 750. Figure~\ref{fig:coating_chamber} shows a picture of the vacuum chamber and associated instrumentation used to coat the ArCLights. The~first step for the production of ArCLights is to laminate the dichroic mirror on a \qtyproduct{32 x 34}{\centi\meter} aluminium plate using a 3M 467MP adhesive \mbox{transfer-tape}.

The aluminium plate is then placed with the dichroic mirror facing downwards on the structure inside the vacuum chamber, such that the dichroic mirror has a distance of \SI{10.5}{\centi\meter} to the heating table. To~keep the sample at room temperature, a~water-cooled plate is attached on top of the aluminium plate. A~tray is filled with uniformly distributed TPB and placed on the custom-made heating table. The~heating table consists of 24 resistive heating elements of type RND 15550 3R9 F mounted on an aluminium plate. The~setup has temperature sensors on the heating table, the~dichroic mirror, and~the water-cooled~plate.

First, the~vacuum chamber is evacuated using a roughing pump and a turbo-molecular pump to start evaporation. The~heating table switches on once the chamber reached a pressure of about \SI{1E-3}{\milli\bar}. The~heating table is connected to a PID controller (ACS-13A-R/M), which keeps the TPB tray temperature constant at \SI{200}{\celsius} by interrupting the~circuit.

\begin{figure}[H]
 \hspace{-6pt}   \includegraphics[trim={0 1cm 0 0.5cm},width=1\textwidth]{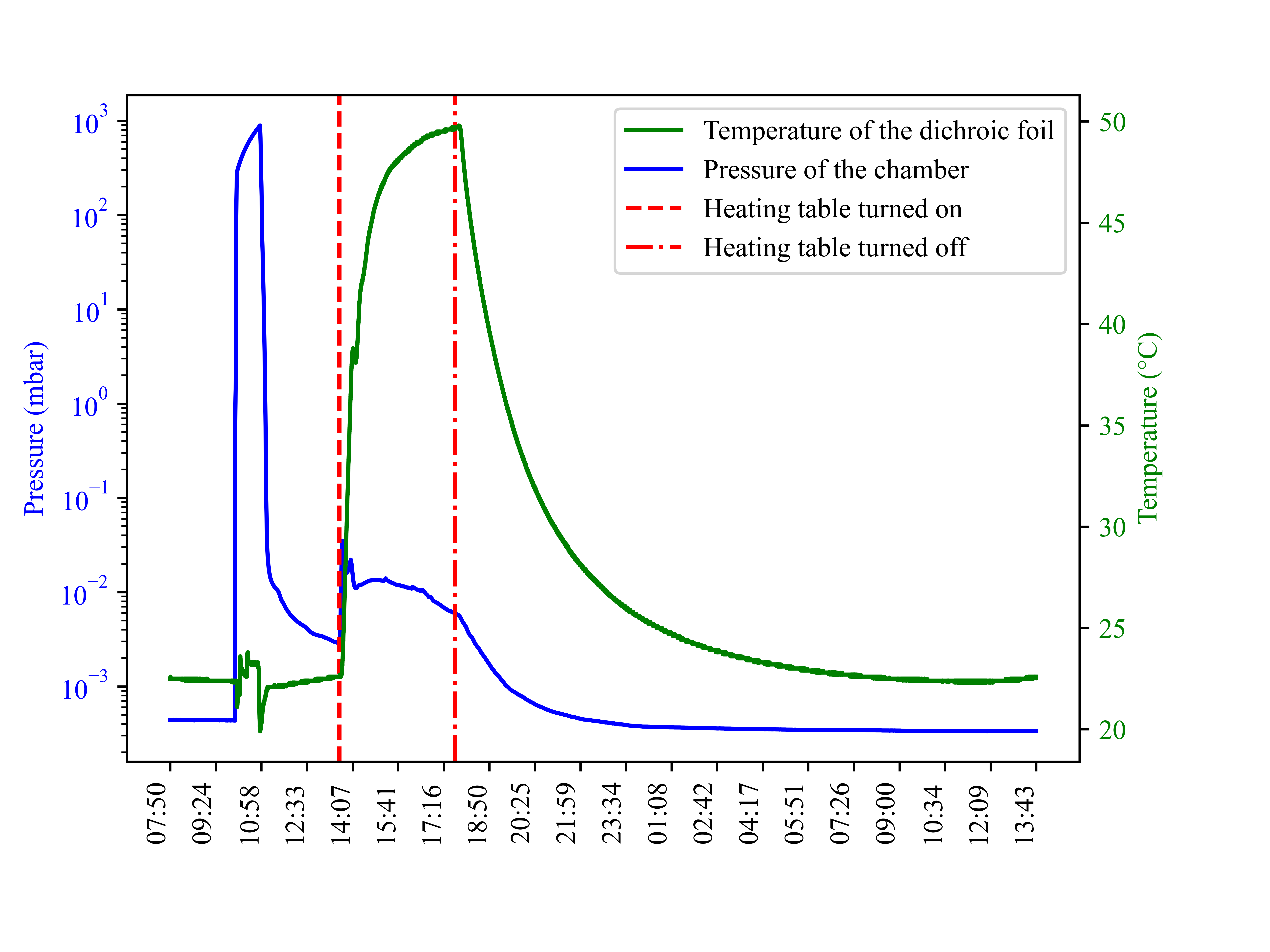}
    \caption{{A} TPB evaporation cycle overview, illustrating the chamber pressure (blue), the~temperatures of the dichroic mirror (green) and the turning on and off of the heating table (red). The~first blue peak represents the moment the chamber is opened to place the dichroic mirror inside. The~pumping process lowers the pressure before the heating table is activated. Once the heater is turned on, water and TPB evaporate, leading to increased pressure within the chamber. The~temperature of the foil increases as the heating table~operates.}
    \label{fig:evaporation_cycle}
\end{figure}

The full coating process, which includes evacuation and heating, takes approximately four hours. Figure~\ref{fig:evaporation_cycle} shows the pressure and temperature of components during the TPB evaporation process.
At the end of the cycle, the~heater is turned off. After~the chamber cools down, the~chamber is pressurised, and~the coated foil is~removed.

\begin{figure}[H]
  \includegraphics[width=0.49\linewidth]{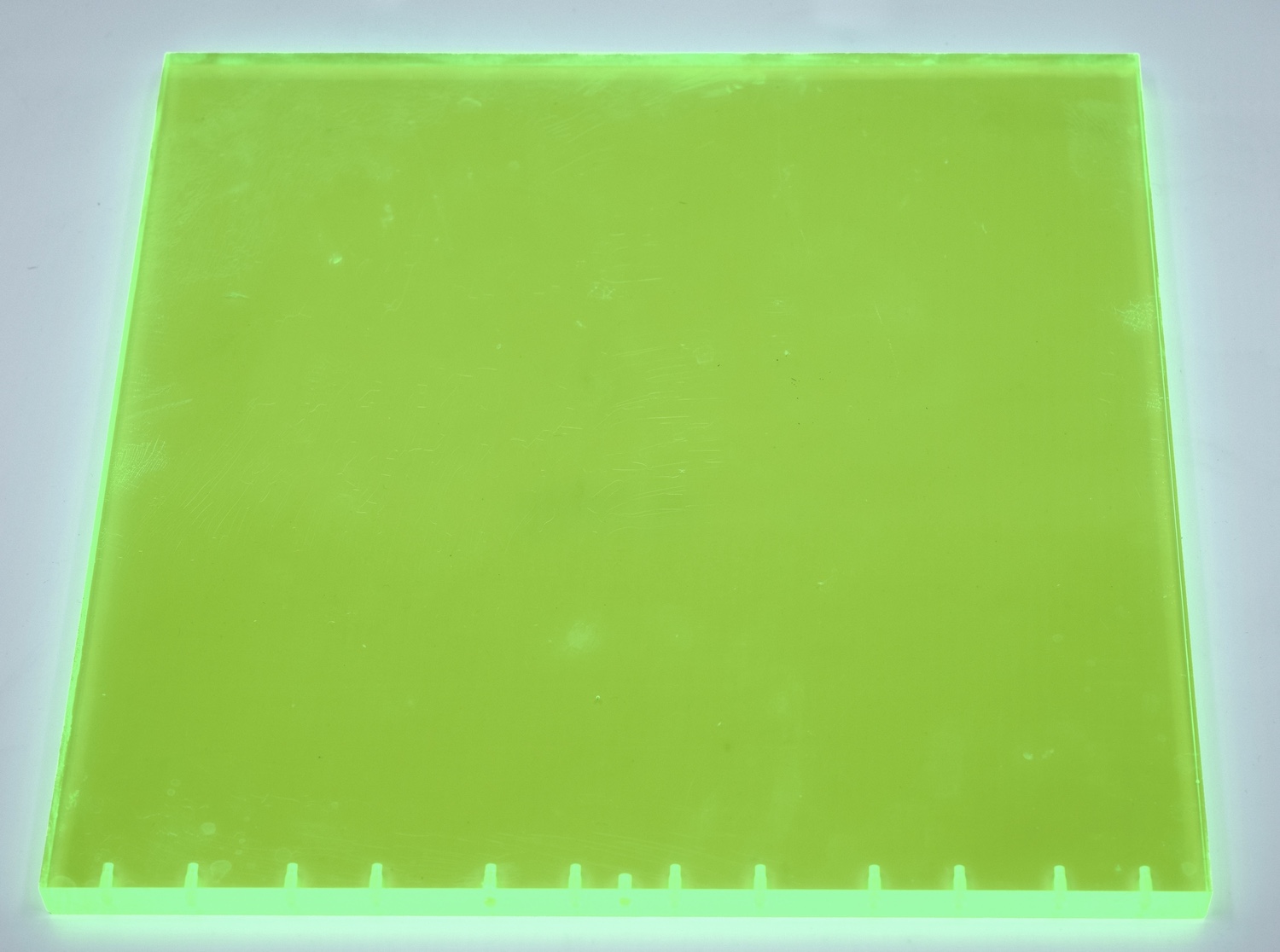}
  \hfill
  \includegraphics[width=0.49\linewidth]{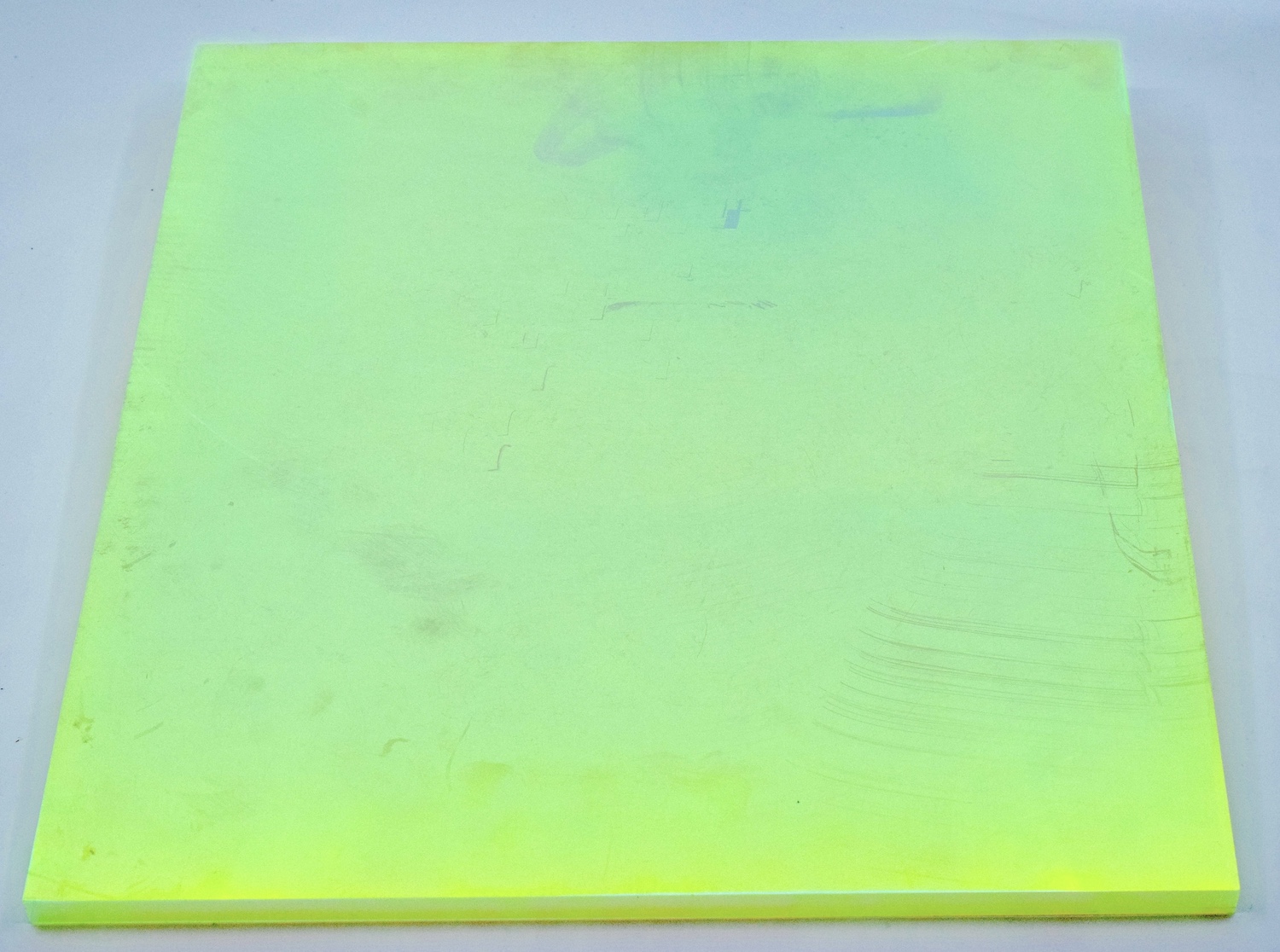}
  \caption{A bare WLS plate EJ280 with prepared threads for SiPM mounting (\textbf{left}). ArCLight after attachment of the TPB-coated foil and edge mirrors (\textbf{right}).}
  \label{fig:finished_acl}
\end{figure}

The coated dichroic mirror is removed from the aluminium plate and glued to the clean WLS EJ280 tile. The~EJ280 material is cleaned with soap and water and dried with a cotton cloth. On~three narrow faces of the EJ280 tile, strips of dichroic mirror are attached. Six SiPMs, placed on a Printed Circuit Board (PCB), are fixed on the remaining narrow face of the ArCLight. Eight-millimetre-long M3 threads are cut directly into the WLS plastic to attach the SiPM PCB with Polyether Ether Ketone (PEEK) screws. This method ensures direct contact between the SiPMs and the light trap. The~area around the SiPMs is covered with a dielectric mirror foil. Figure~\ref{fig:finished_acl} shows a bare WLS tile EJ280 on the left and a WLS tile with TPB-coated foil on the~right.

\vspace{-6pt}

\section{Quality Control and Analysis of Produced~ArCLights} \label{sec:qaqc}

Twenty-two units were produced following the procedure described in Section~\ref{sec:production}. Once the ArCLight is produced, two quality assurance tests are conducted to ensure its performance. The~first test is a qualitative optical inspection. The~coating structure and TPB crystals are observed with an optical microscope (VHX-7000, {Keyence, Osaka, Japan}).

\begin{figure}[H]
  \includegraphics[width=0.85\textwidth]{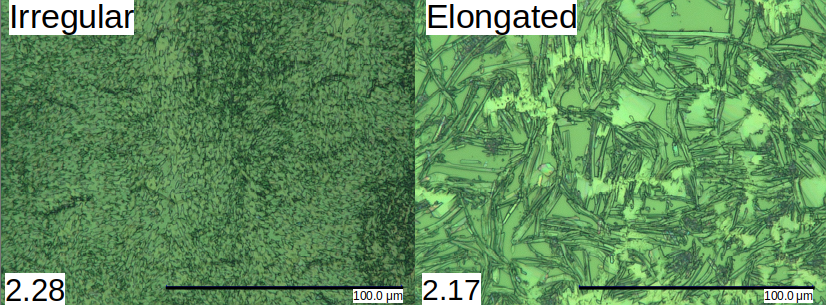}
  \caption{{Example} of the two different crystals observed on the coated ArCLight. The~magnification of both images is the~same.}
  \label{fig:crystal_def}
\end{figure}

Two different types of crystals of TPB, referred to as irregular and elongated, are identified. Figure~\ref{fig:crystal_def} shows an example of each. The~irregular crystals are small pad-like crystals that create an irregular distribution over the area. The~elongated crystals are filament-like crystals that are bunch-wise~oriented.
 
\begin{figure}[H]
    \includegraphics[width=0.68\textwidth]{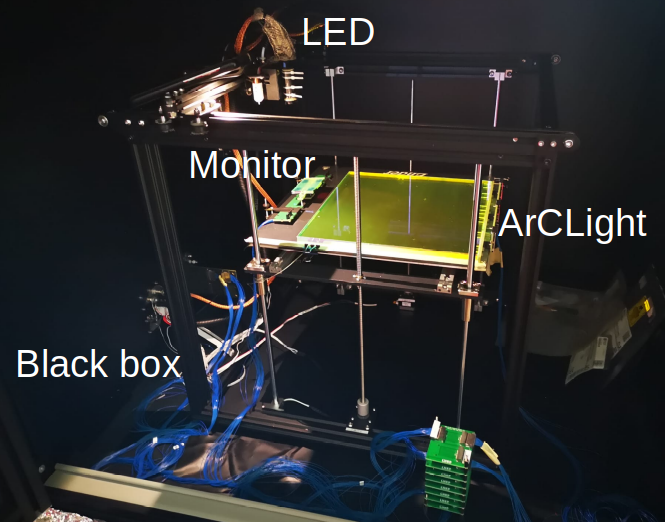}
    \hfill
    \caption{The black box used to scan ArCLights. The~movable LED that produces light during scanning is shown on the~top.}
    \label{fig:scaner_concept}
\end{figure}

\begin{figure}[H]
    \includegraphics[width=0.65\textwidth]{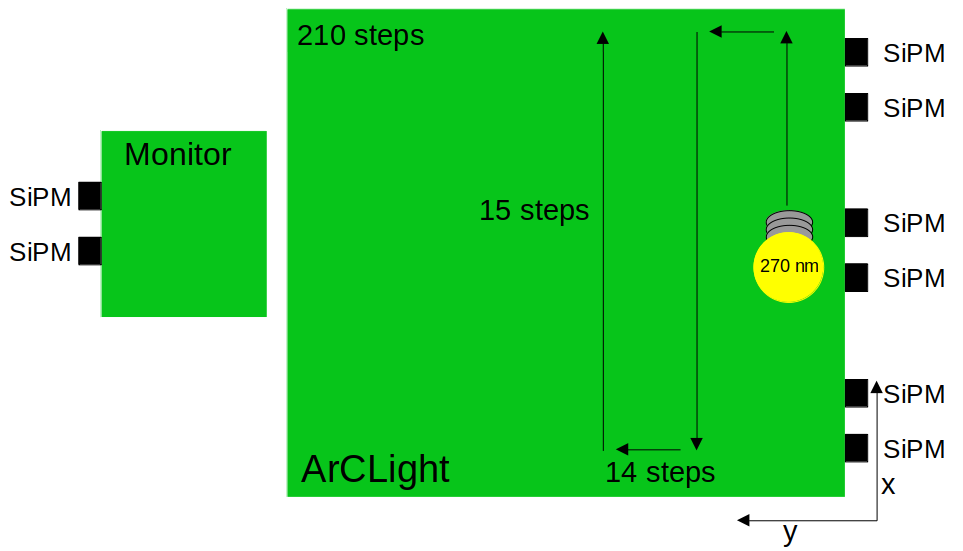}
    \caption{Conceptual sketch of the scanning procedure of an~ArCLight.}
    \label{fig:scaner_concept2}
\end{figure}

The second test quantitatively compares the light yield for all ArCLights in the production batch. This test uses a pulsed LED that emits light with a wavelength of 
\SI{270}{\nano\meter}. The~LED is mounted onto a robotic arm that can move across the surface of the ArCLight. The~test measures the light yield at 210 different LED positions to scan the ability of light collection across the surface of the ArCLight. The~LED is kept at a fixed vertical distance from the ArCLight surface during the scan. The~measurement is performed in a black box, which shields the setup from stray light. The~black box and the scanning setup are shown in Figure~\ref{fig:scaner_concept}. In~Figure~\ref{fig:scaner_concept2}, a~cartoon of the scanning process is~sketched. 

\vspace{-6pt}

\begin{figure}[H]
\centering
\begin{subfigure}{.5\textwidth}
  \centering
  \includegraphics[width=1.0\linewidth]{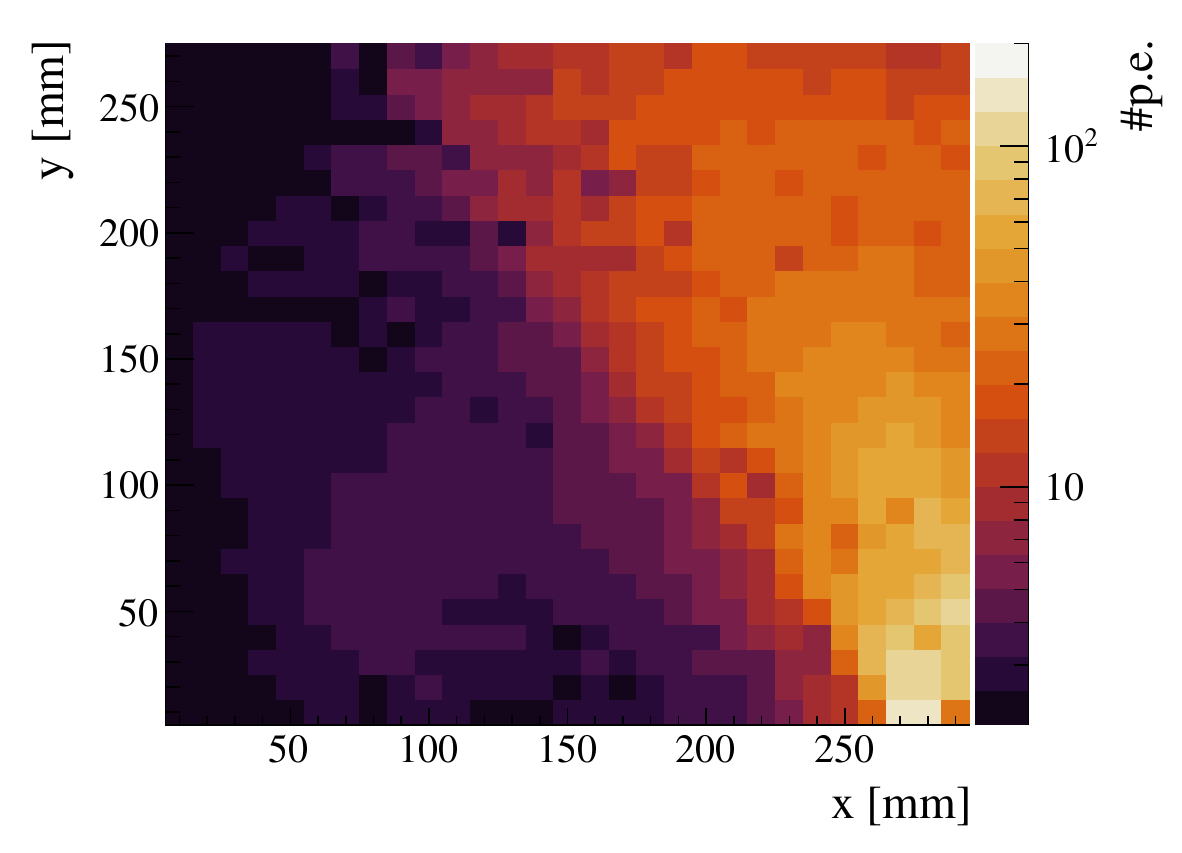}
  \caption{SiPM~0}
  \label{fig:nph0}
\end{subfigure}%
\hfill
\medskip
\begin{subfigure}{.5\textwidth}
  \centering
  \includegraphics[width=1.0\linewidth]{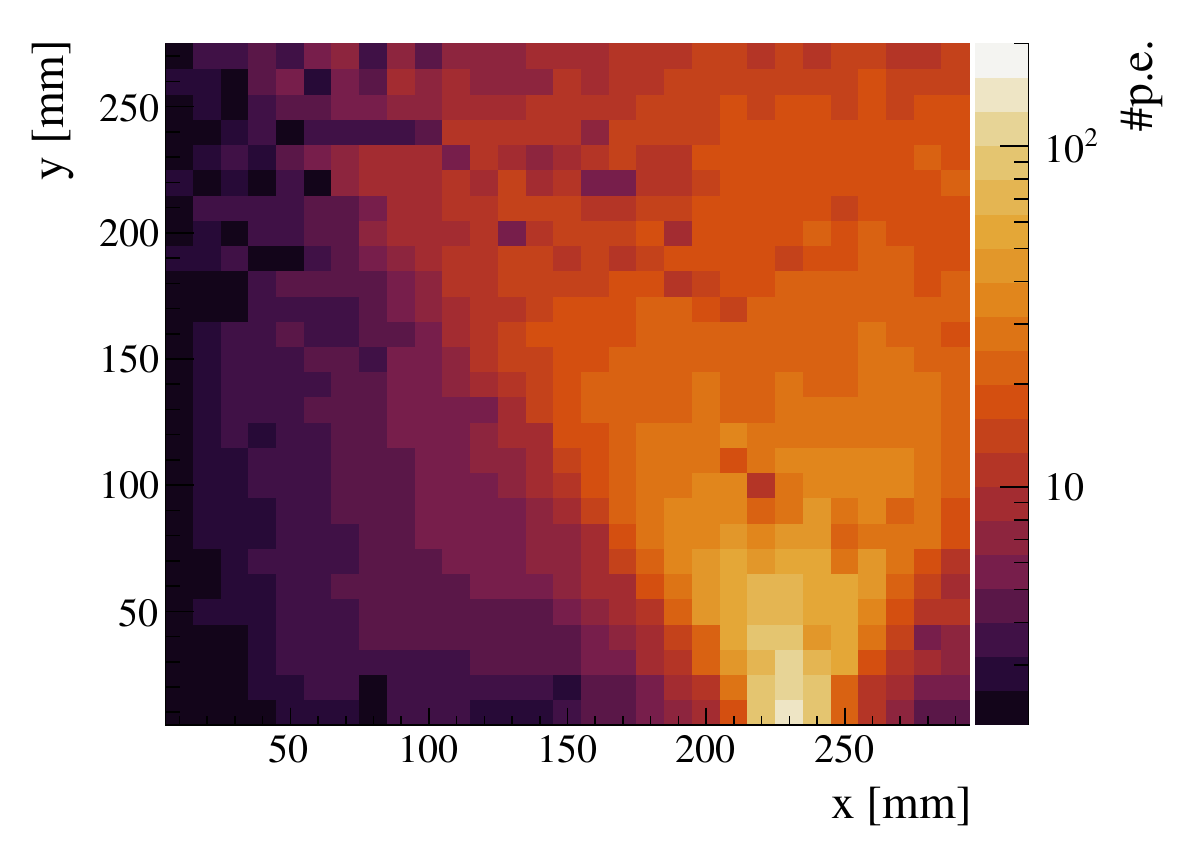}
  \caption{SiPM~1}
  \label{fig:nph1}
\end{subfigure}%
\hfill
\medskip
\begin{subfigure}{.5\textwidth}
  \centering
  \includegraphics[width=1.0\linewidth]{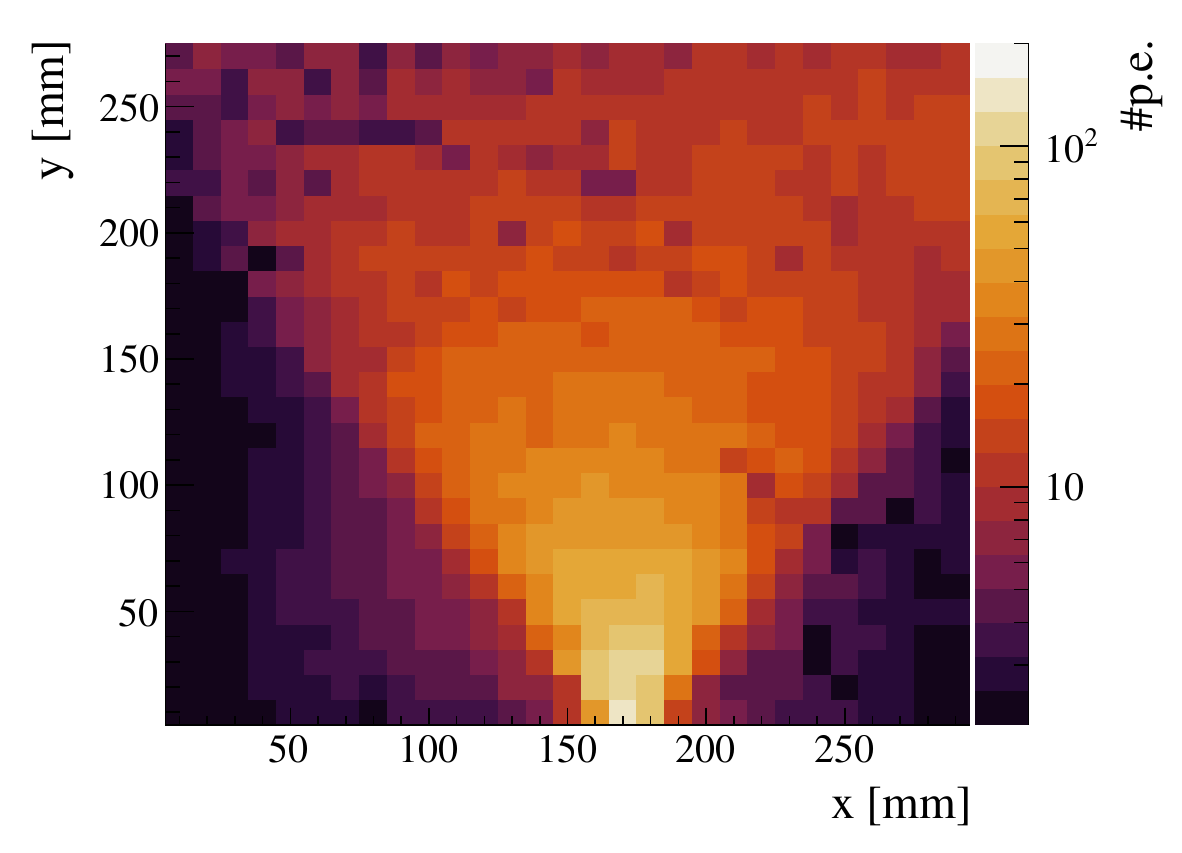}
  \caption{SiPM~2}
  \label{fig:nph2}
\end{subfigure}%
\hfill
\medskip
\begin{subfigure}{.5\textwidth}
  \centering
  \includegraphics[width=1.0\linewidth]{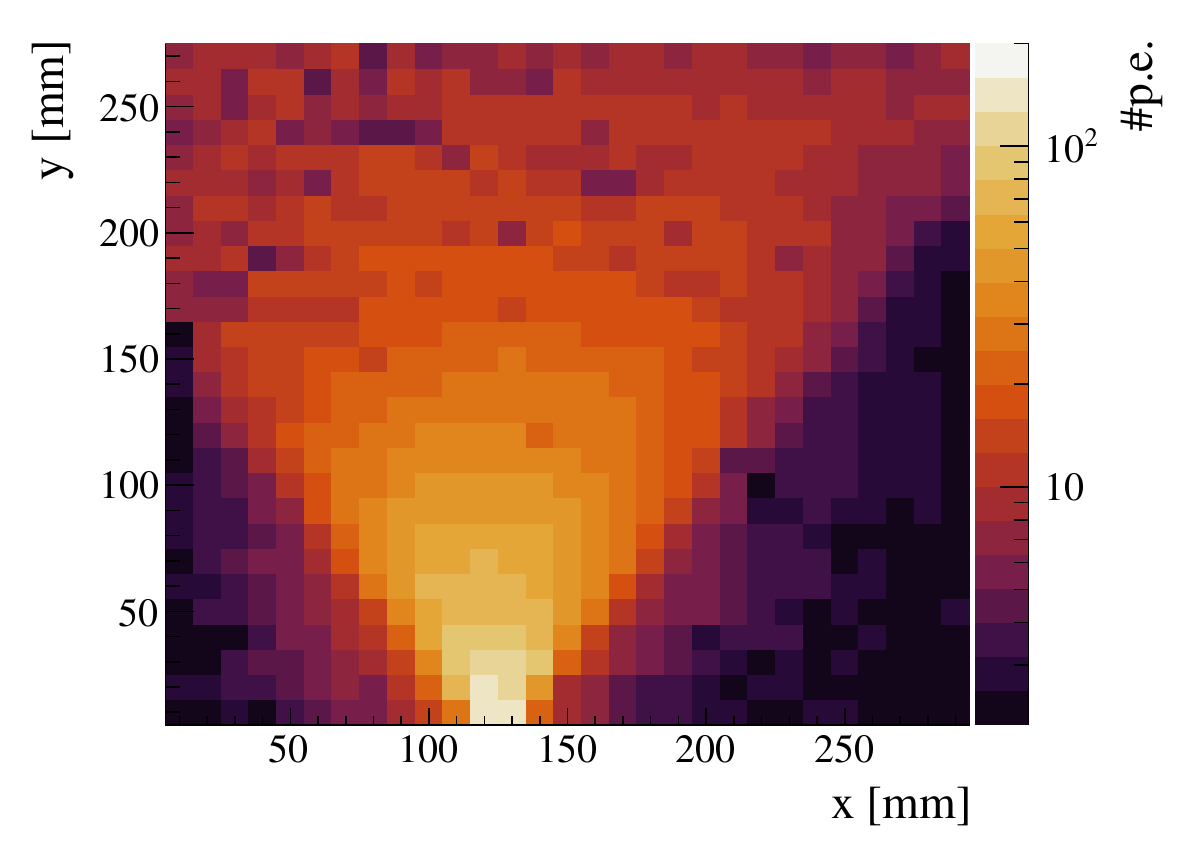}
  \caption{SiPM~3}
  \label{fig:nph3}
\end{subfigure}%
\hfill
\medskip
\begin{subfigure}{.5\textwidth}
  \centering
  \includegraphics[width=1.0\linewidth]{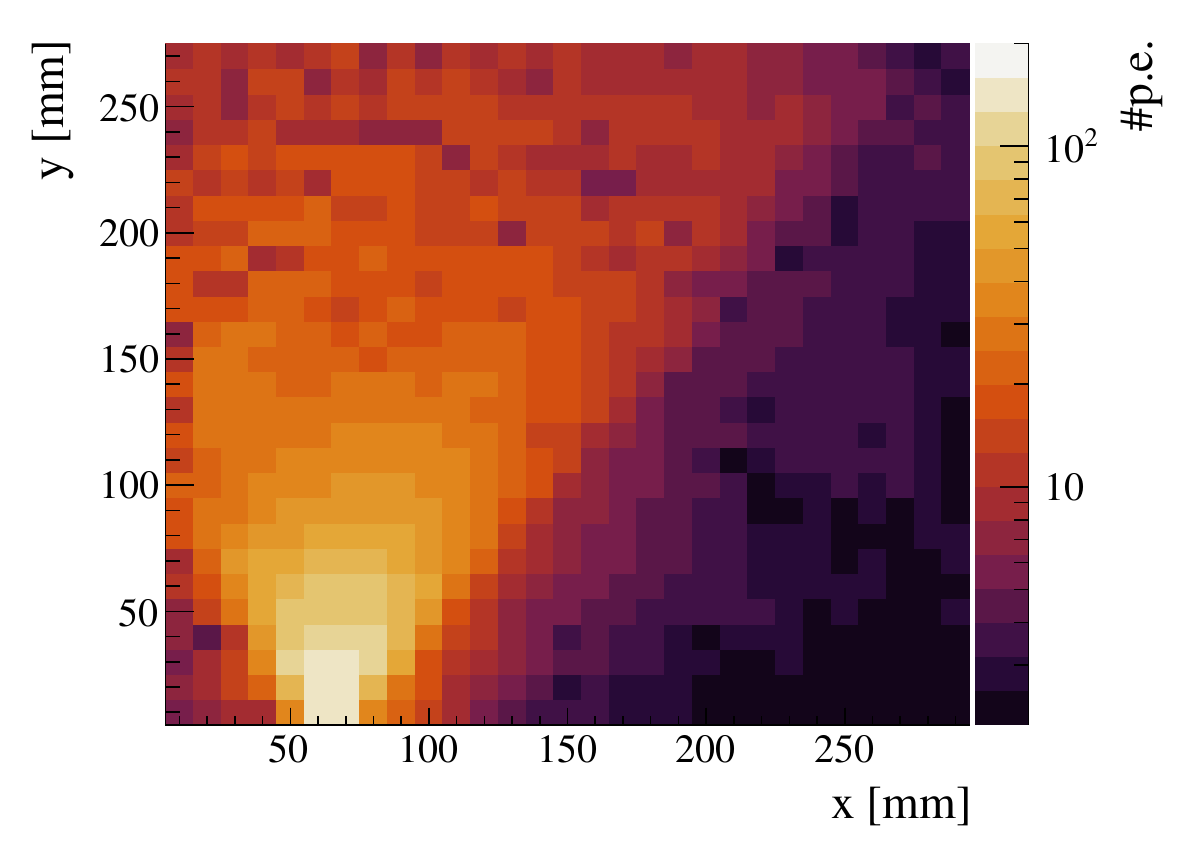}
  \caption{SiPM~4}
  \label{fig:nph4}
\end{subfigure}%
\hfill
\medskip
\begin{subfigure}{.5\textwidth}
  \centering
  \includegraphics[width=1.0\linewidth]{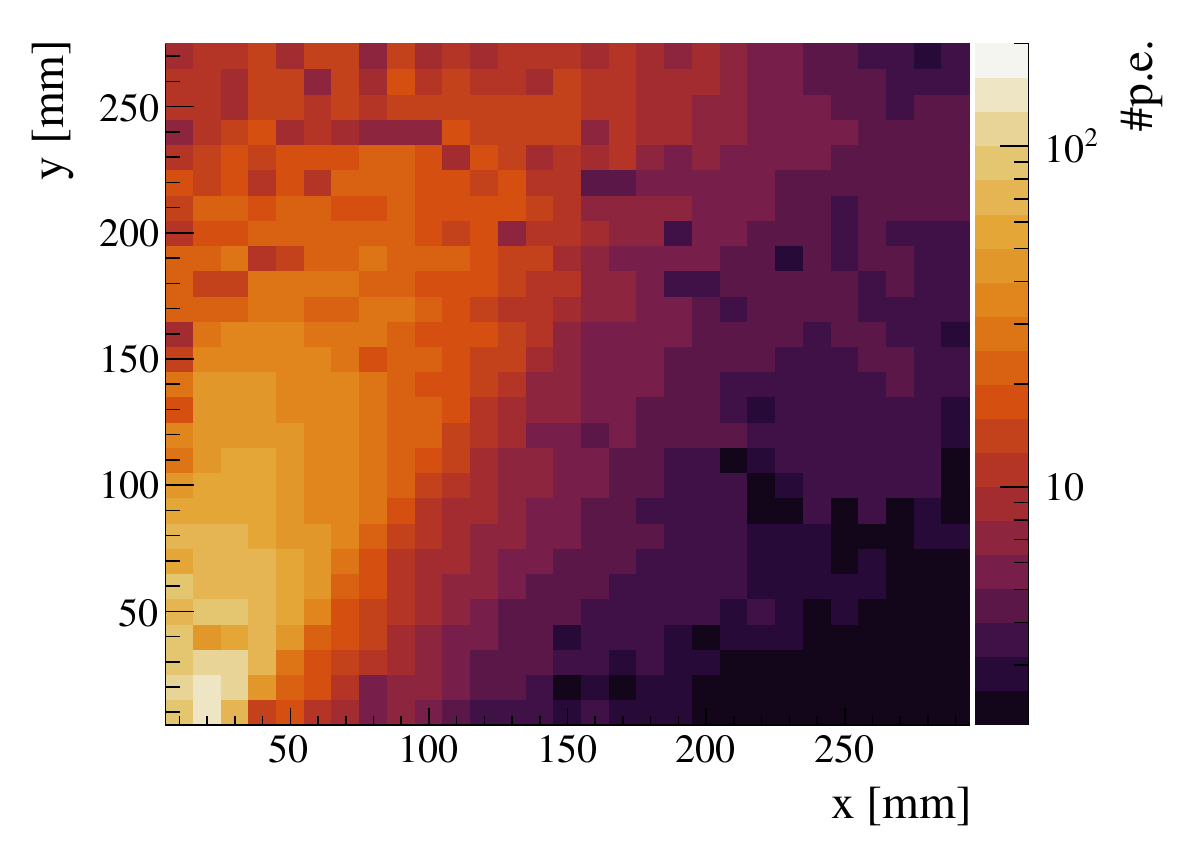}
  \caption{SiPM~5}
  \label{fig:nph5}
\end{subfigure}%
\caption{The detected number of {p.e.s} in a high-resolution scan by each of the six SiPM channels for the LED at a distance of \SI{20}{\milli\meter}, with~the SiPMs positioned on the x-axis. The~colour scale represents the detected number of photoelectrons {(p.e.s).} The~absolute signal strength is arbitrary and depends on the LED light emission power, which is kept constant for the different~scans.}
\label{fig:sing_sipm_scan}
\end{figure}

The LED pulse triggers the light readout system, which records the waveforms from the six SiPMs. The~waveforms are integrated over the pulse width and calibrated by measuring response waveforms of single photoelectrons {(p.e.s)}. 
For~each position of the LED and on each SiPM, 15,000 waveforms are collected, and~the mean number of p.e. per SiPM is extracted. Figure~\ref{fig:sing_sipm_scan} shows the result of a high-resolution~scan. 

\begin{figure}[H]
\hspace{-0.7cm}  \includegraphics[trim={0 0.5cm 0 0}, width=0.9 \textwidth]{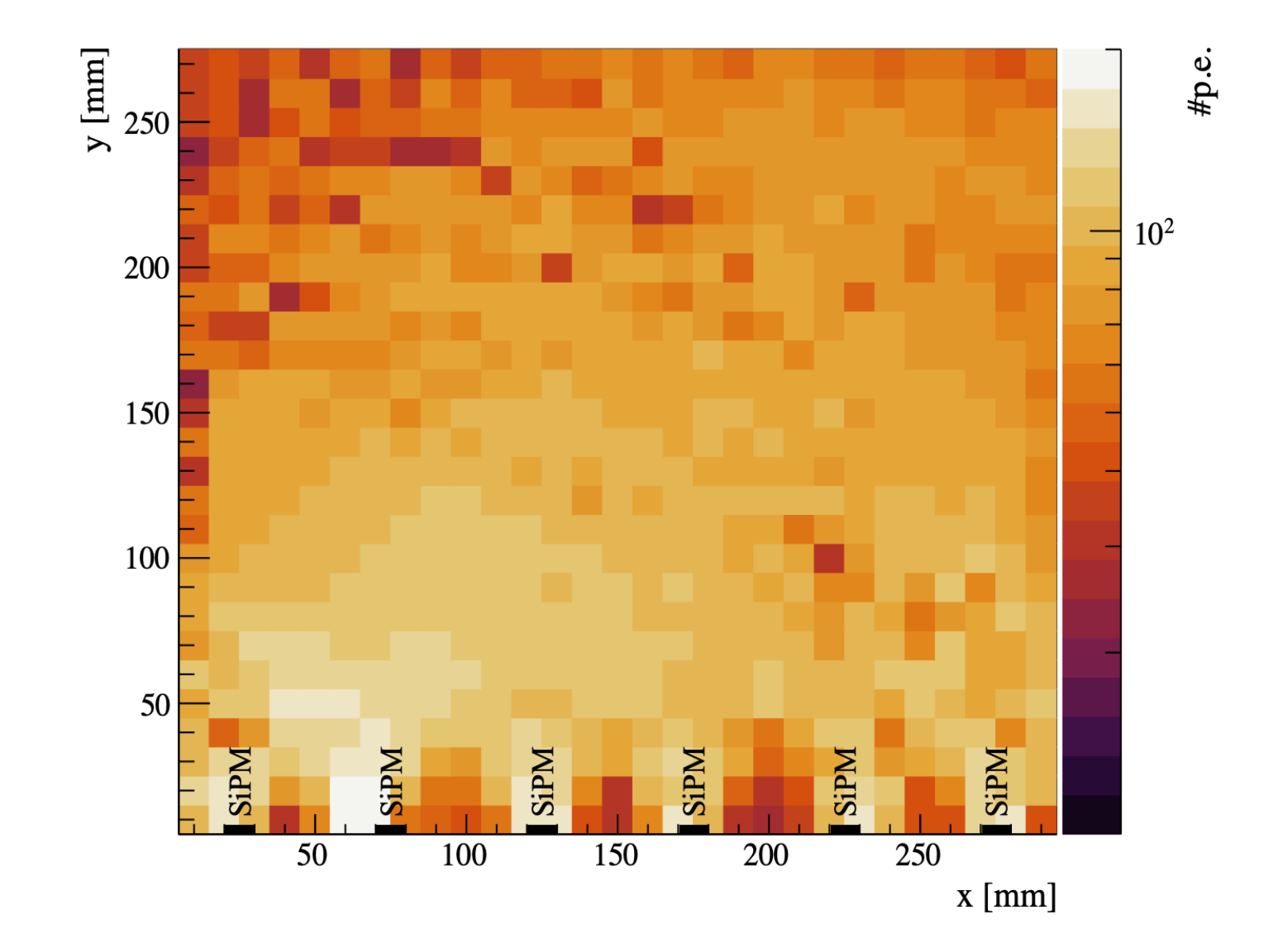}
    \caption{Example scan of an ArCLight with corrected light yields using monitoring SiPMs. The~scan measures the total photoelectrons of 27 $\times$ 29~positions.}
    \label{fig:scan_tile_2_11}
\end{figure}

For a performance comparison, the~mean of the collected{ p.e.s} of the six SiPMs are summed for each position of the LED. The~sums of the collected light per position illustrates the number of captured {p.e.s} for each SiPM at every position. As~shown in Figure~\ref{fig:scan_tile_2_11}, the~amount of the collected light mainly depends on the distance to the SiPMs. Additionally, observed reductions in light yield could indicate regions of poor TPB coverage. \mbox{Two additional} SiPMs fixed on the scanning table monitor the LED and measure its light yield stability. 
They enable us to correct for intensity variations. During~a scan, the~monitoring SiPMs are repeatedly illuminated by the LED. The~measured number of {p.e.s} is equalised and the number of {p.e.s} on the ArCLight SiPMs is subsequently adjusted for each~point.\vspace{-12pt}

\begin{figure}[H]
\hspace{-0.7cm} \includegraphics[width=\textwidth]{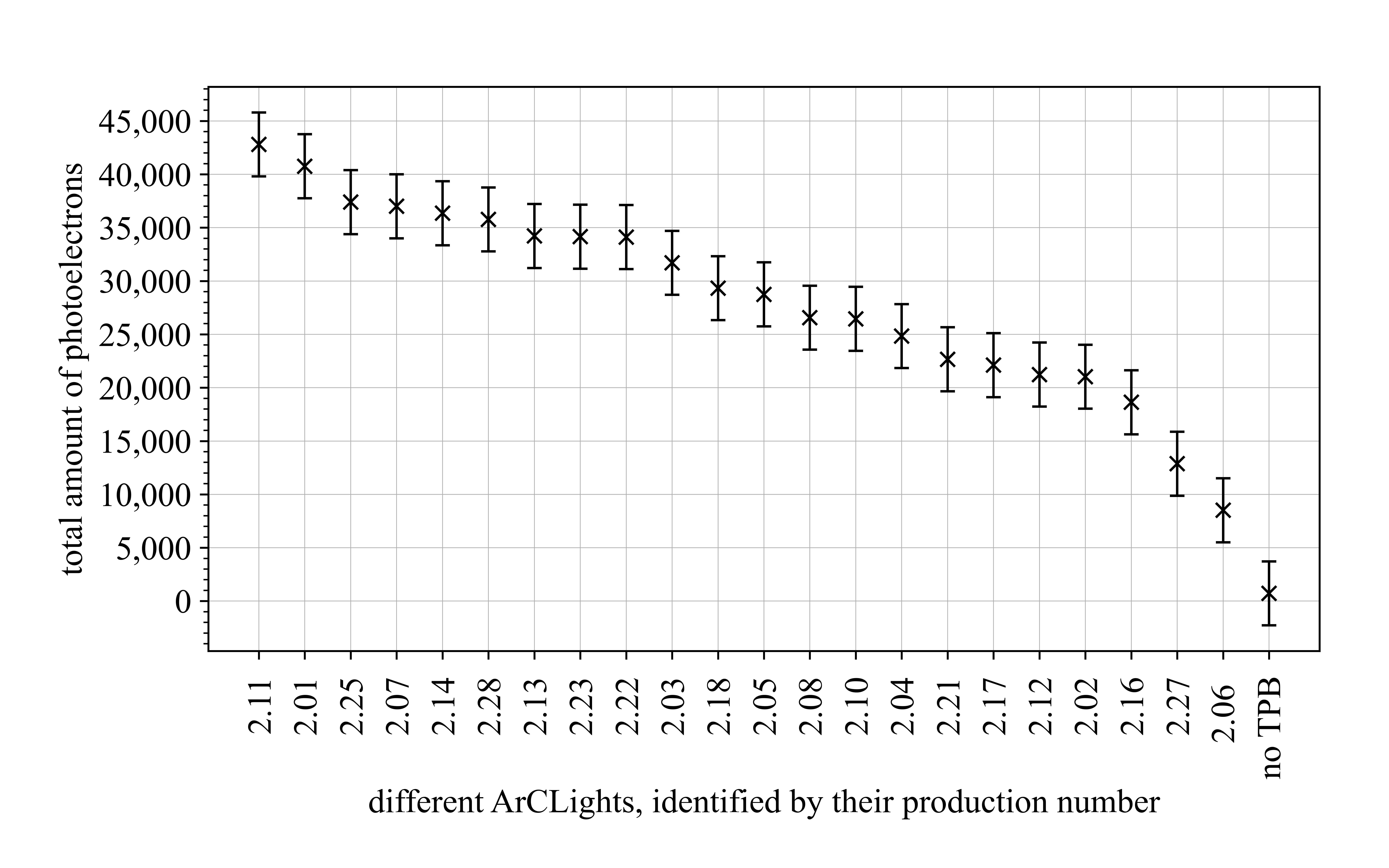}\vspace{-9pt}
    \caption{{For} each ArCLight, the~means of the collected {p.e.s} for every position are added up to a total amount of {p.e.s} collected per scan. The~different performances of the ArCLight are shown based on the total {p.e.s} measured per scan. The~error bars are estimated based on the largest difference observed between scans of the same~ArCLight.}
    \label{fig:Light_sum}
\end{figure}

The total light yield across all positions scanned on the ArCLight is used as a metric to compare different ArCLights. The~comparison of the different ArCLights produced in the production batch for the ArgonCube 2 $\times$ 2 Demonstrator is shown in Figure~\ref{fig:Light_sum}. The~uncertainties on the measurements are set at 3000 {p.e.s}, a~conservative estimate based on the maximum disagreement observed between scans of the same ArCLight. The~ArCLight with the worst light collection is the ArCLight without any TPB coating, which {is} 
analysed for comparison. This demonstrates the necessity of  TPB in order to detect any photons in the UV range. Figure~\ref{fig:Light_sum} shows a large spread of performance for different ArCLights, highlighting the importance of stabilising and understanding the production parameters as much as possible in future productions. Specific differences in the ratio of irregular to elongated crystals covering the ArCLights are~observed.

Twenty ArCLights were selected with similar and consistent TPB coverage and inspected with a microscope. For~each ArCLight, at~least five randomly distributed \qtyproduct{150 x 180}{\micro\meter} surface regions were evaluated with respect to their predominant crystal shapes. These were then categorised by the fraction of elongated and irregular crystals observed. While no ArClight was found to have a majority of elongated crystals, \mbox{two categorisations} emerged: ArCLights with a ratio of 1:1 irregular to elongated crystals, and~ArCLights with a ratio of over 2:1 irregular to elongated crystals. For~each ArClight, the~total amount of {p.e.s} was determined using the scanning technique described earlier in this section. The~average of the total light collected for the two categories revealed an increase of $(36 \pm 8)\%$ in light yield for the sample of ArCLights with a higher fraction of irregular crystals.

\section{Conclusions}

Significant advancements have been made to enhance the performance and design of the ArCLight system. The~transition from manually brushing the TPB layer onto the surface to evaporating it within a vacuum chamber has notably increased the coverage and uniformity of TPB distribution. This change improved the photon detection efficiency by approximately a factor of~two. 

The performance variations for different ArCLights in the scan suggest a potential instability in production conditions. Improvements are planned in the control of the sample cooling and TPB heating during the coating procedure in future productions. The~scans show that the TPB layer and the crystal shape are crucial factors for the performance of the ArCLights. Nevertheless, quality assurance assessments confirm that ArCLights meet the performance requirements for their intended use in DUNE ND-LAr and the \mbox{2 $\times$ 2}~Demonstrator.

No significant changes in techniques are anticipated for the final 50 cm tile. The~collection efficiency decreases with increasing tile length, and~the 50 cm tiles are expected to exhibit a slightly lower overall efficiency. While this is not expected to be detrimental, further improvements are envisioned to maximize the detector performance of the ArCLight~technology.

\vspace{6pt} 


\authorcontributions{{Methodology}, L.C., F.F., A.G., I.K., J.K. and S.P.; Formal analysis, L.C., F.F., A.G., I.K., J.K. and S.P.; Investigation, J.B., L.C., F.F., A.G., L.F.I., J.K. and S.P.; Writing---original draft, L.C., R.D., A.G. and J.K.; Writing---review and editing, L.C., R.D., J.K and M.W.; Supervision, I.K. and M.W. All authors have read and agreed to the published version of the manuscript.} 

\funding{{The research} is funded by grant 200021-169045 of Swiss National Science Foundation and by Canton of Bern, Switzerland. This~study is supported by the Mechanic and Electric Workshop at the Laboratory for High Energy Physics, Universität Bern, as~well as the Albert Einstein Centre for Fundamental Physics, Universität~Bern.} 

\dataavailability{The original contributions presented in the study are included in the article, further inquiries can be directed to the corresponding author.}


\conflictsofinterest{The authors declare no conflicts of~interest.} 



\appendixtitles{no} 
\appendixstart

\begin{adjustwidth}{-\extralength}{0cm}

\reftitle{References}



\PublishersNote{}
\end{adjustwidth}
\end{document}